\documentclass[onecolumn, 10pt,english,aps]{revtex4}
\usepackage[T1]{fontenc}
\usepackage{array}
\usepackage{graphicx}
\usepackage[all,dvips]{xy}
\usepackage{amssymb}

\usepackage{babel}

\begin{document}

\title{Perpetual points and periodic perpetual loci in maps}

\author{Dawid Dudkowski$^1$, Awadhesh Prasad$^2$, and Tomasz Kapitaniak$^1$}
\affiliation{$^{1}$Division of Dynamics, Technical University of Lodz, Stefanowskiego 1/15, 90--924 Lodz, Poland}
\affiliation{$^{2}$Department of Physics and Astrophysics, University of Delhi, Delhi 110007, India}

\begin{abstract}
We introduce the concepts of perpetual points and periodic perpetual loci in discrete--time systems (maps). The occurrence and analysis of these points/loci are shown and basic examples are considered. We discuss the potential usage and properties of introduced concepts. The comparison of perpetual points and loci in discrete--time and continuous--time systems is presented. Discussed methods can be widely applied in other dynamical systems.
\vspace{\baselineskip}

\end{abstract}

\maketitle

\textbf{Perpetual points, which have been inspired by the theory of stationary points, are the natural extention of the well--known equilibrium states of the system. They have been widely observed in nonlinear dynamical models, and their study has provided many useful applications and properties of these points. Although the concept has been originally described in systems of ODE's, it can be effectively introduced also in different types of models, e.g. discrete--time systems. In our paper we define perpetual points in the maps and extend the concept into periodic perpetual loci, which can be naturally compared with the standard periodic orbits observed in the system. The analysis of the occurence and the properties of perpetual points and loci in most elementary, as well as more complex examples of maps is presented. Additionally, we have compared the proposed concepts in the systems described by discrete--time and continuous--time equations, exhibiting potential similarities and differences between them. The obtained results expand our knowledge of the concepts of perpetual points and loci and their potential usage in dynamical systems.}
\vspace{\baselineskip}

\textbf{1. Introduction}
\vspace{\baselineskip}

Perpetual points are a new type of critical points in dynamical systems introduced by Prasad in \cite{pp1}. 
They are defined as points for which acceleration of the system becomes zero while velocity remains nonzero
 and can be found widely in literature \cite{pp1,pp2,pp3,pp4,pp5,pp6}. The topological conjugacy for these points
is discussed in \cite{pp3}.  There are various interesting properties of those points. They can 
be used to understand various behavior of the systems \cite{pp1,pp2}.

Although perpetual points are still quite a new idea in the dynamical systems theory, a significant amount of research needs to be conducted, if we want to understand the issue, ideas of new connections, and its applications properly \cite{pp6}.
The existence of these points also confirms whether a system is dissipative or not \cite{pp1} (dissipativity in the sense of Levinson, see e.g. \cite{diss}). However,
this confirmation has limitation for specific type of  systems, see \cite{pp4}.
It has also been found that  these points are important for better understanding of transient dynamics in the phase
space \cite{pp1}.

Many real--world problems, e.g.  population growth \cite{may}, compound interest \cite{fin}, radioactive decay 
\cite{radio}, medication dosages \cite{dose}, alternate bearing \cite{bearing1,bearing2} etc. are modeled using discrete--time systems. 
Although, studies on dynamics of maps are less popular than for ODE's, 
research on this kind of systems is still ongoing and many new results are observed. Discrete--time switched linear 
systems \cite{maps2}, multi--agent systems \cite{maps4} or finite--time control \cite{maps3} are only a few 
examples of such analysis.  Considering the importance of studying  such  systems, in this paper, we
introduce the  concept of perpetual point  in discrete--time systems.
Note that in the original paper \cite{pp1}, the existence and applications of perpetual points are discussed only in
 continuous--time systems.

Very recently, a new type of attractors called hidden attractors, that don't intersect with the neighborhood
of any fixed point  have been a topic of discussion  \cite{hidden1,pp5,hidden3,hidden4,hidden5,hidden6,hidden7,hidden8}.
It can be found in many systems, e.g.  Chua's circuit \cite{hidden4}, van der Pol--Duffing
\cite{hidden5},  radio--physical oscillators \cite{hidden6}, exponential nonlinear terms \cite{hidden8} 
as well as in coupled systems \cite{hidden7,pp2}.
Due to the absence of unstable fixed point in its neighborhood these type of attractors are less
tractable and hence  it is also difficult to understand their characteristic properties.
Existence of perpetual points lead to the hypothesis, that they can be a natural guidance to the hidden 
attractors \cite{hidden1,pp5,hidden3,hidden4,hidden5,hidden6,hidden7,hidden8}, analogically to the fixed  
points which locate self--excited attractors (details are presented in \cite{pp2}).
This connection have been also thoroughly analyzed in \cite{pp5}.
 Recently, hidden attractors  have been observed also in discrete--time
systems \cite{hiddenmaps1,hiddenmaps2,hiddenmaps3,hiddenmaps4}. This new observations and the fact, that 
perpetual points are possibly connected to hidden attractors have been a motivation to introduce an analogy of 
these points in maps in this paper.

The paper is organized as follows. In Section 2 we introduce the concepts of perpetual points and periodic perpetual loci in maps and present examples in typical systems. In Section 3 the comparison of perpetual points/loci in discrete--time and continuous--time systems is shown. The summary of the obtained results is presented in Section 4.
\vspace{\baselineskip}

\textbf{2. Results}
\vspace{\baselineskip}

\textbf{2a. Concept of perpetual points in discrete--time systems}
\vspace{\baselineskip}

Let us consider the $n$--dimensional discrete--time system given by equations:
\begin{equation}
x^{i}_{t+1} = f_{i} (x^{1}_{t}, \ldots, x^{n}_{t}),
\label{map}
\end{equation}
where $i=1, \ldots, n$ and $t \in \mathbb{N} \cup \left\{ 0 \right\}$ is discrete time. Here, $(x^{1}_{t}, \ldots, x^{n}_{t})$ denotes the state of the system in time $t$.

Based on the definition of derivative for continuous functions (as the limit of difference quotient), we introduce analogous mathematical structure for the discrete case. Let $d: S \rightarrow S$, where $S$ is a sequence space (e.g., $S=l^{p}$) be an operator defined as:
\begin{equation}
d x^{i}_{t} = \frac{x^{i}_{t+1}-x^{i}_{t}}{t+1-t} = x^{i}_{t+1}-x^{i}_{t} = f_{i} (x^{1}_{t}, \ldots, x^{n}_{t}) - x^{i}_{t},
\label{d-der}
\end{equation}
where $i \in  \left\{ 1, \ldots, n \right\}$ is a considered space variable. We will call sequence $d x^{i}$ the \textit{discrete derivative} of sequence $x^{i}$.

The fixed point $(x^{1}_{fp}, \ldots, x^{n}_{fp})$ of the system (\ref{map}), is the one for which the condition $f_{i} (x^{1}_{fp}, \ldots, x^{n}_{fp}) = x^{i}_{fp}$ is satisfied for every $i \in  \left\{ 1, \ldots, n \right\}$. As we can see, it is the same point for which the \textit{discrete derivative} $d x^{i}_{fp} = 0, i \in  \left\{ 1, \ldots, n \right\}$. This represents an analogy to the continuous--time dynamical systems, for which the derivative of each space variable equals zero at equilibrium.

Using the definition (\ref{d-der}) of the operator $d$ we can consider the \textit{higher order discrete derivatives} as the consecutive compositions of operator $d$ with itself. Hence, let:
\begin{equation}
d^{2} x^{i}_{t} = \frac{d x^{i}_{t+1} - d x^{i}_{t}}{t+1-t} = x^{i}_{t+2} - 2 x^{i}_{t+1} + x^{i}_{t} = f_{i} (f_{1} (x^{1}_{t}, \ldots, x^{n}_{t}), \ldots, f_{n} (x^{1}_{t}, \ldots, x^{n}_{t})) - 2 f_{i} (x^{1}_{t}, \ldots, x^{n}_{t}) + x^{i}_{t}
\label{dd-der}
\end{equation}
be a \textit{second order discrete derivative} of space variable $x^{i}$.

As an analogy to \textit{perpetual points} in continuous--time dynamical systems, we can introduce similar concept in maps using relation (\ref{dd-der}). Namely, we will call the point $(x^{1}_{pp}, \ldots, x^{n}_{pp})$ for which $d^{2} x^{i}_{pp} = 0 \wedge  d x^{i}_{pp} \neq 0, i=1, \ldots, n$ the \textit{perpetual point} of system (\ref{map}).
\vspace{\baselineskip}

\textbf{2b. Examples in typical maps}
\vspace{\baselineskip}

(i) Logistic map
\vspace{\baselineskip}

Let us consider a typical example of one--dimensional discrete--time dynamical system, the
logistic map \cite{log1}. Although the system is very well--known, it still inspires researchers and new results can be observed in many areas of science (e.g. in encryption algorithms \cite{log2,log3}).

The dynamics is given by equation:
\begin{equation}
x_{t+1} = a x_{t} (1-x_{t}),
\label{log}
\end{equation}
where $a \in (0,4]$ is the system parameter. Relation (\ref{log}) transforms interval [0,1] into itself and we consider the dynamics only in this interval.

\begin{figure}
\includegraphics[scale=0.55]{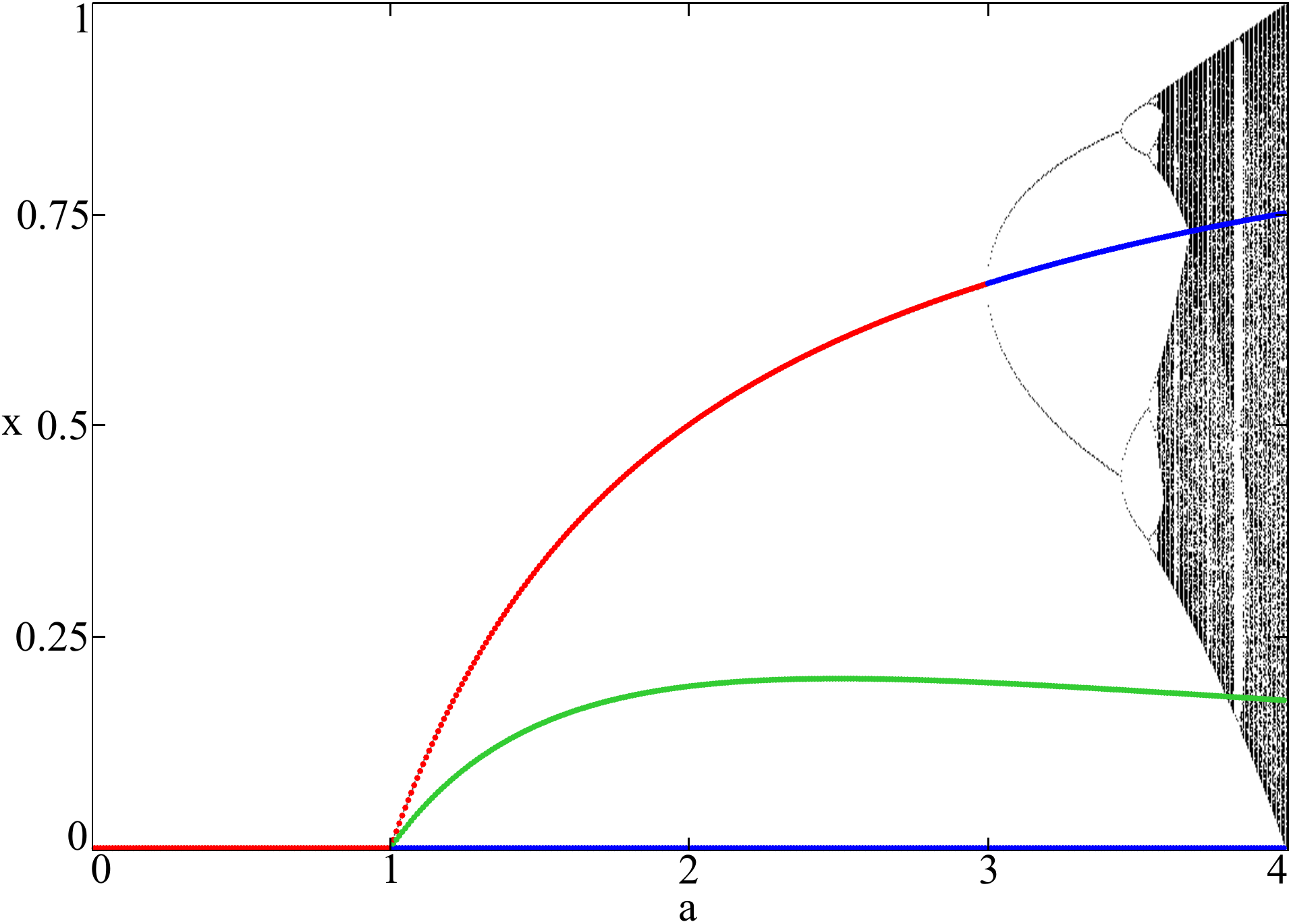}
\caption{(color online). The bifurcation diagram of system (\ref{log}) (black dots) for parameter $a \in (0,4]$. Stable fixed points are shown as red dots, while unstable equilibria are presented as blue ones. Green dots correspond to perpetual points of the system.}
\label{fig1}
\end{figure}

A typical bifurcation scenario and critical points of the system (\ref{log}) are shown in Fig.~\ref{fig1}.

For increasing parameter $a$, a typical road to chaos is shown as the background of Fig.~\ref{fig1} (black dots). One can observe a sequence of period doubling bifurcations leading to the chaotic behavior for $a \gtrsim 3.56$. The stable and unstable fixed points are shown by red and blue dots respectively. The zero equilibrium $x=0$ is stable for $a \in (0,1]$ and loses stability when $a=1$. The second fixed point $x=(a-1)/a$ is born for $a=1$ and attracts all the trajectories till the first period doubling bifurcation at $a=3$. For $a \in (1,4]$ system (\ref{log}) has one unique perpetual point marked in Fig.~\ref{fig1} by green dots. As one can see, the perpetual points appear along with stable fixed points through the bifurcation that occurs at $a=1$. Moreover, both of the critical points are born near the same point of state space, i.e. at $x=0$.

Discrete--time systems can exhibit different types of dynamics even in their simplest form. In a one--dimensional map one can observe not only equilibria, but also periodic orbits and chaotic behavior (while in case of continuous--time systems such states require the dimension of at least two and three respectively). To compare perpetual points with more complex attractors, we consider the system:
\begin{equation}
x_{t+1} = f^{k} (x_t),
\label{k-log}
\end{equation}
where $f(x)=a x (1-x)$ describes the dynamics of logistic map and $k \in \mathbb{N}$ denotes the number of compositions of function $f$ with itself. Fixed points of system (\ref{k-log}) are period-$k$ orbits of system (\ref{log}). Similarly, we will call perpetual points of (\ref{k-log}) the \textit{period-$k$ perpetual loci} of map (\ref{log}) (the definition can be easily applied to general equation (\ref{map})).

\begin{figure}
\includegraphics[scale=0.55]{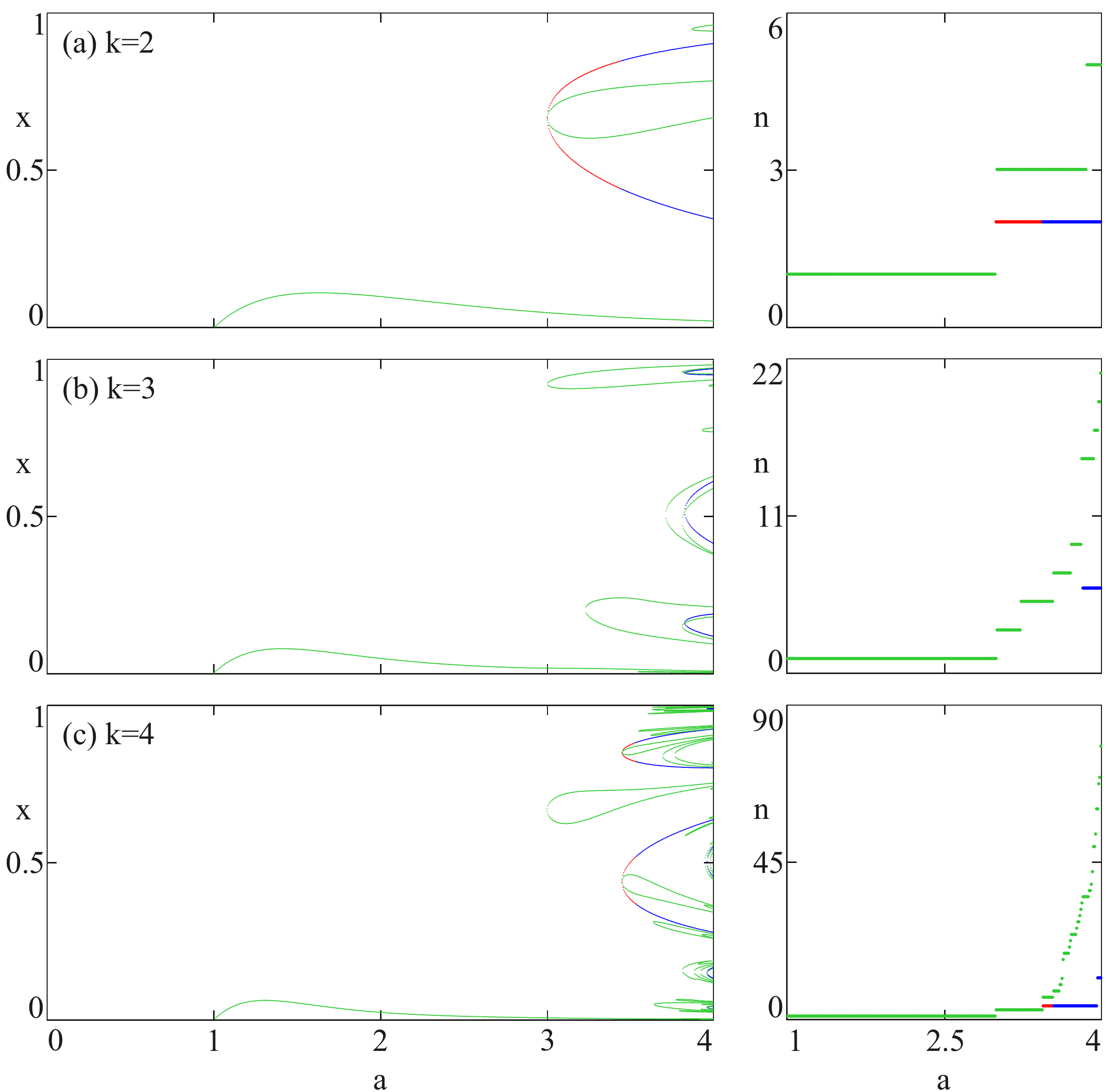}
\caption{(color online). In the left panel the periodic orbits (red -- stable, blue -- unstable) and perpetual loci (green) are shown for period $k=2, 3, 4$ (increasing from the top to the bottom). Number of co--existing points ($n$ parameter) is presented in the right panel for $a \in (1,4]$ (case $n=0$ have been omitted).}
\label{fig1k}
\end{figure}

In Fig.~\ref{fig1k} results of our calculations for system (\ref{k-log}) are shown. In the left panel one can observe the values of both critical points for $a \in (0,4]$ (color code as in Fig.~\ref{fig1}, i.e. red/blue dots denote stable/unstable orbits and green ones correspond to perpetual loci). The periods $k=2, 3, 4$ have been considered ((a), (b), (c) respectively). The co--existence of periodic loci can be observed for $a>3$ (for $a \in [1,3]$ only unique perpetual loci are present). There are a few scenarios in which perpetual loci can be born. They can appear around standard bifurcations of the system (\ref{log}), i.e. at the saddle--node bifurcation (at $a=1$) or at period doubling ones, when two branches of perpetual loci are created along with standard periodic orbits (e.g. at $a=3$ for $k=2$). On the other hand, this type of loci appear for many other different parameter values, for which logistic map does not exhibit any qualitative changes. Moreover, it seems that all the way they are created pairwise. Results suggest, that the route of periodic perpetual loci can be more complex than corresponding periodic attractors.

As can be seen in Fig.~\ref{fig1k}, number of periodic perpetual loci increase along with increase of parameter $a$ value, and their position in phase space get more complex. In the right panel of Fig.~\ref{fig1k} one can observe the number $n$ of these loci (green) compared to periodic orbits (stable -- red, unstable -- blue) for $a \in (1,4]$. For $k=2$ one, three or five perpetual loci exist depending on the parameter, but if we increase the period $k$, more of these loci are born (with the maximum of 21 and 87 for $k=3$ and $k=4$ respectively). Those values are much higher then the number of co--existing periodic attractors. What should be noted, this growth increase rapidly near parameter region where chaotic behavior occurs ($a \gtrsim 3.56$).

\begin{figure}
\includegraphics[scale=0.55]{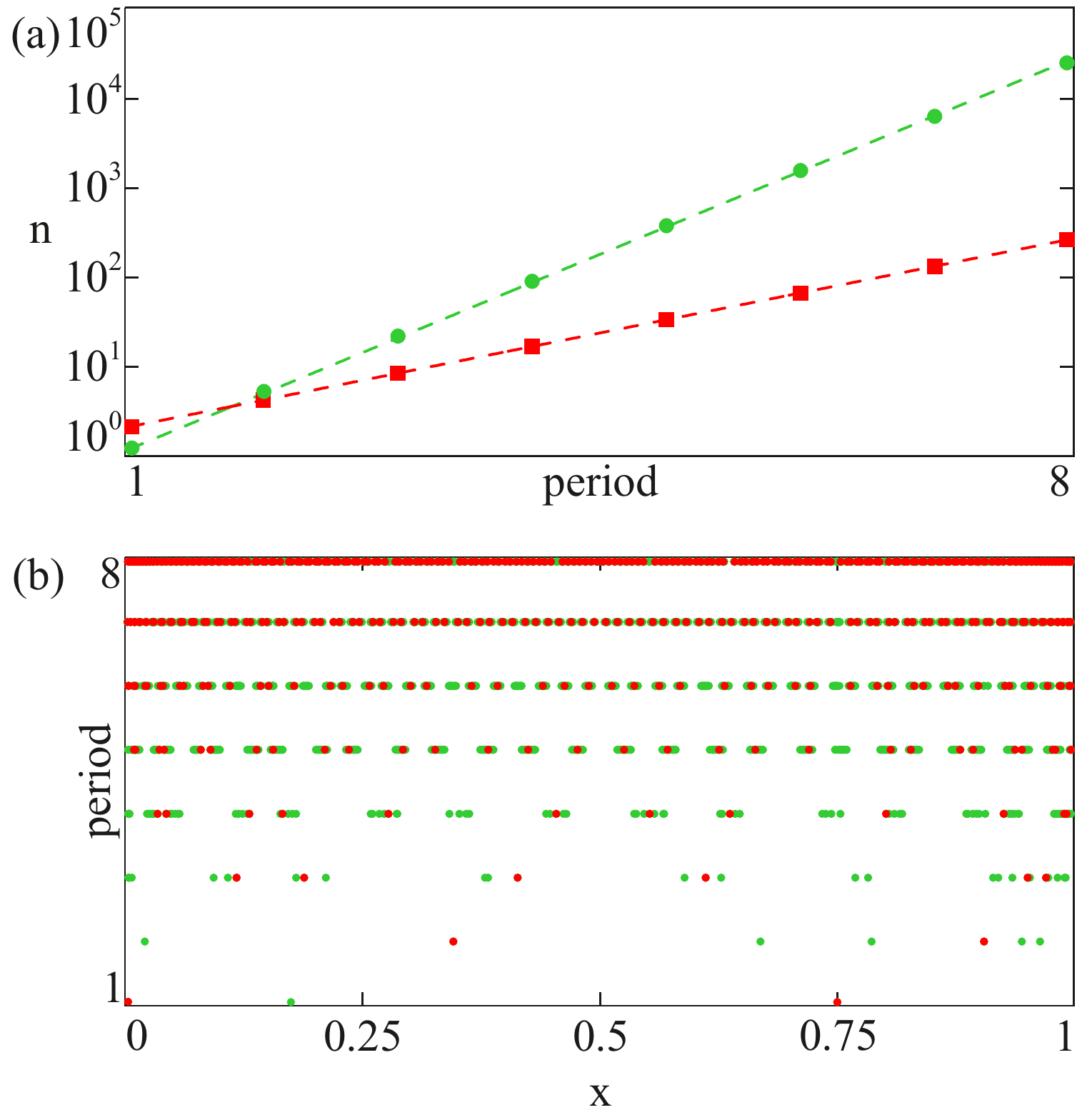}
\caption{(color online). In (a) the number of co--existing periodic orbits (red squares) and perpetual loci (green circles) is presented for periods equal $1, \ldots, 8$  (scale on vertical axis is logarithmic). The fitting lines are shown as red (periodic) and green (perpetual) dashed curves. In (b) the position in phase space of both types of states is shown.}
\label{fig1stat}
\end{figure}

In order to investigate this phenomenon, we have fixed the parameter $a$ and compared the numbers of co--existing periodic orbits and perpetual loci for different values of period \cite{scaling}. In Fig.~\ref{fig1stat} results obtained for $a=4$ are presented. In Fig.~\ref{fig1stat}(a) the number $n$ of perpetual loci (green circles) and periodic states (red squares) are shown for period value ranging from one to eight (horizontal axis). It should be emphasized that the scale on vertical axis is logarithmic. Number of periodic orbits (points that belong to these orbits) equals $2^{k}$, where $k$ is the period (we count all the states which period does not exceed $k$ value), and the number of corresponding perpetual loci exceeds this value significantly with increase of $k$. We have fitted the latter results using least squares method for exponential functions, i.e. $p_1 \cdot {p_2}^{k}$, where $p_1, p_2>0$ are positive parameters (operator '$\cdot$' denotes typical multiplication in real numbers). The number of obtained points have been approximated as $0.2672 \cdot {4.213}^{k}$. The fitting lines for both types of states are shown in Fig.~\ref{fig1stat}(a) as dashed lines (where red and green curves correspond to periodic orbits and perpetual loci respectively).

Based on our calculations, the ratio between periodic perpetual loci and unstable periodic orbits for $a=4$ equals $0.2672 \cdot {2.1065}^{k}$. Consequently, the set of former states is much more denser than the latter ones. This have been presented in Fig.~\ref{fig1stat}(b), where the position in phase space of both sets of states is marked for fixed period on vertical axis (color code as in Fig.~\ref{fig1stat}(a)).
\vspace{\baselineskip}

(ii) Bi--stable tent map
\vspace{\baselineskip}

As the second example, we consider the bi--stable tent map \cite{tent2,tent3} (studies on the original mono--stable system can be found in \cite{tent1}), which is given by:
\begin{eqnarray}
   x_{t+1} = f(x_{t}) = \left\{
     \begin{array}{ll}
     p x_{t} + ({p}/{l}-1) & : x_{t} \in [-1, -{1}/{l})\\
     l x_{t} & : x_{t} \in [-{1}/{l}, {1}/{l})\\
     p x_{t} - ({p}/{l}-1) & : x_{t} \in [{1}/{l}, 1],
     \end{array}
   \right.
	\label{tent}
\end{eqnarray}
where $l$ and $p$ are the parameters and depending on their values \cite{tent2} system transforms interval $[-1,1]$ into itself. Equivalently, the dynamics can be described by $f(x_{t}) = p x_{t} + \frac{l-p}{2} ( \left| x_{t} + 1/l \right| - \left| x_{t} - 1/l \right| )$.

\begin{figure}
\includegraphics[scale=0.55]{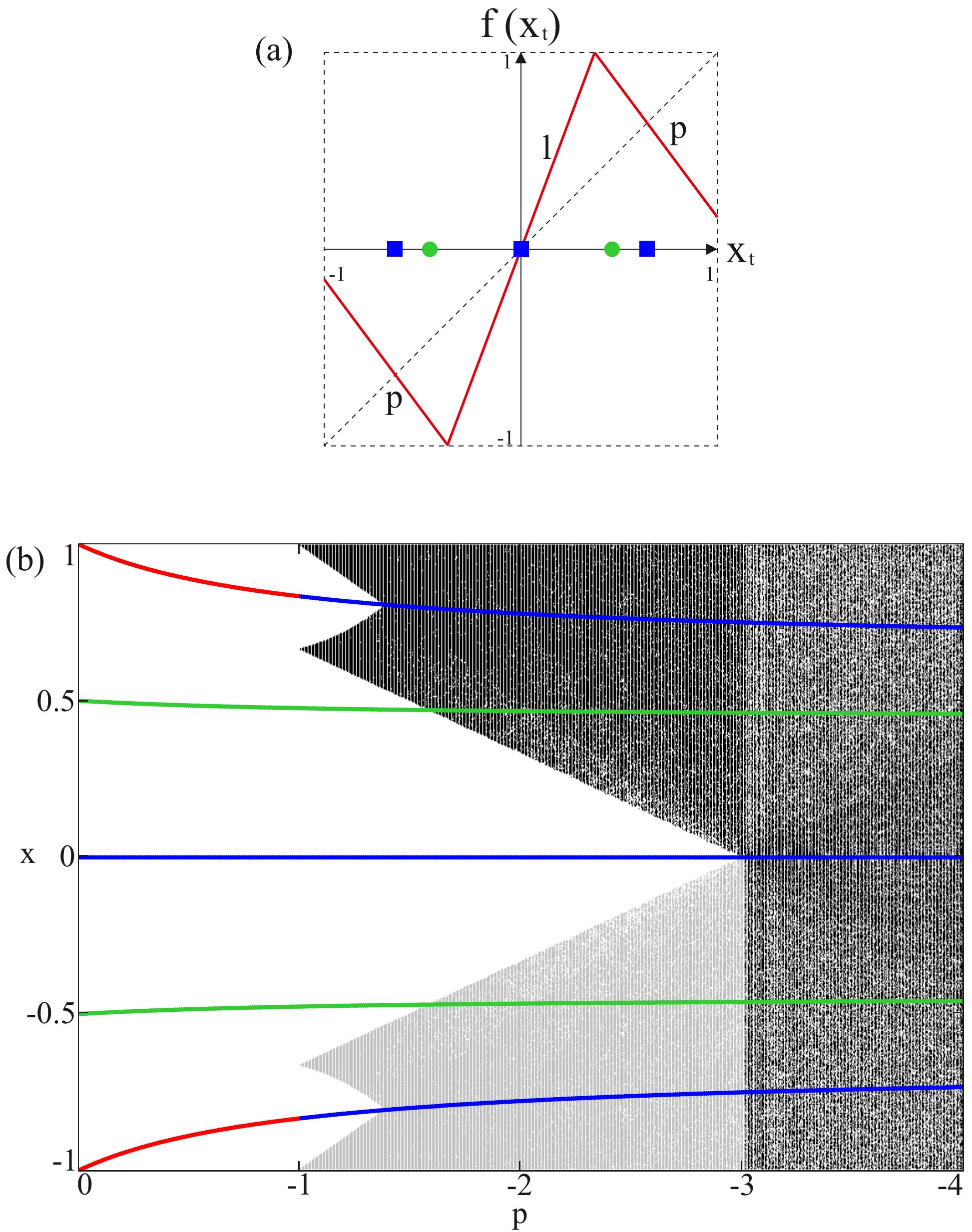}
\caption{(color online). In (a) the local dynamics of system (\ref{tent}) described by function $f$ is presented, while in (b) the bifurcation diagram for parameter $p \in [-4, 0)$ is shown. Co--existing attractors are marked in black and grey. In both subfigures equilibria are marked in red and blue (stable and unstable respectively), while perpetual points are denoted in green.}
\label{fig2}
\end{figure}

In Fig.~\ref{fig2}(a) the local dynamics of function $f$ is shown. As one can see, it consists of three linear parts, where parameters $l$ and $p$ describe their slopes (here, $l=1.5$ and $p=-2.4$). In \cite{tent2} it has been shown, that depending on the parameters values, the system (\ref{tent}) can be mono--stable having one global chaotic attractor (as the whole $[-1,1]$ interval) or it can be bi--stable, having two co--existing attractors -- fixed points or chaotic (also with co--existing $[-1,0)$ and $(0,1]$ basins of attraction).

The system (\ref{tent}) has three fixed points $x=0$ and $x=\frac{\pm(l-p)}{l(p-1)}$ (blue squares in Fig.~\ref{fig2}(a)), while it also has two perpetual points (green dots).

In Fig.~\ref{fig2}(b) the bifurcation scenario for fixed $l=1.5$ and $p \in [-4, 0)$ considered as the bifurcation parameter has been presented (note that parameter $p$ has been decreased from $p=0$ to $p=-4$). The color code corresponds to the previous example of a logistic map. For $p \in [-1, 0)$ two stable fixed points co--exist (red dots). At $p=-1$ critical points lose their stability (blue dots) and two chaotic attractors are born (black and grey dots). The states co--exist for $p \in (-3, -1)$ and at $p=-3$ they combine into one global attractor. The equilibrium $x=0$ is unstable in the whole bifurcation interval.

The perpetual points of the system (\ref{tent}) are denoted by green dots in Fig.~\ref{fig2}(b) and they exist for every $p \in [-4, 0)$ (two co--existing points). For a fixed $p$ value, first of the points exists in interval $[-1,0)$, while the second one in interval $(0,1]$. Thus, in the case of bi--stability, both attractors can be located using perpetual points.

The statistical analysis of bi--stable tent map is as follows. For a fixed parameter $p=-4$ (for which the system is monostable with $[-1,1]$ chaotic attractor) the number of periodic orbits (points belonging to these orbits) equals $2 \cdot 2^{k}$, where $k$ is considered period (the equilibrium $x=0$ have been omitted). On the other hand, the number of perpetual loci have been approximated using similar methods as in previous example (i) and equals $0.5402 \cdot {4.7784}^{k}$. Consequently, the ratio between periodic perpetual loci and unstable periodic orbits
is $0.2701 \cdot {2.3892}^{k}$. This result is close to the one obtained for the logistic map (\ref{log}) (i.e., $0.2672 \cdot {2.1065}^{k}$). Such scaling similarities may result from the fact that the logistic and the tent maps are naturally connected (topologically conjugated). Therefore, in systems of similar dynamics the properties of perpetual points and loci may be universal.
\vspace{\baselineskip}

(iii) Henon map
\vspace{\baselineskip}

Perpetual points can be found not only in simple one--dimensional maps, but also in multi--dimensional systems. As an example we present the results obtained for $2D$ Henon map \cite{hiddenmaps4,henon1,henon2,henon3}. The system is described by the following set of equations:
\begin{eqnarray}
   \left\{
     \begin{array}{l}
     x_{t+1}=1-a x^{2}_{t} + y_{t},\\
		 y_{t+1}=-b x_{t},
     \end{array}
   \right.
	\label{henon}
\end{eqnarray}
 where $(x_t, y_t)$ denotes the position on plane in time $t$, while $a$ and $b$ are the parameters. A thorough analysis of such defined single Henon map (\ref{henon}), as well as the dynamics of coupled system of such maps can be found in \cite{henon4}.

\begin{figure}
\includegraphics[scale=0.7]{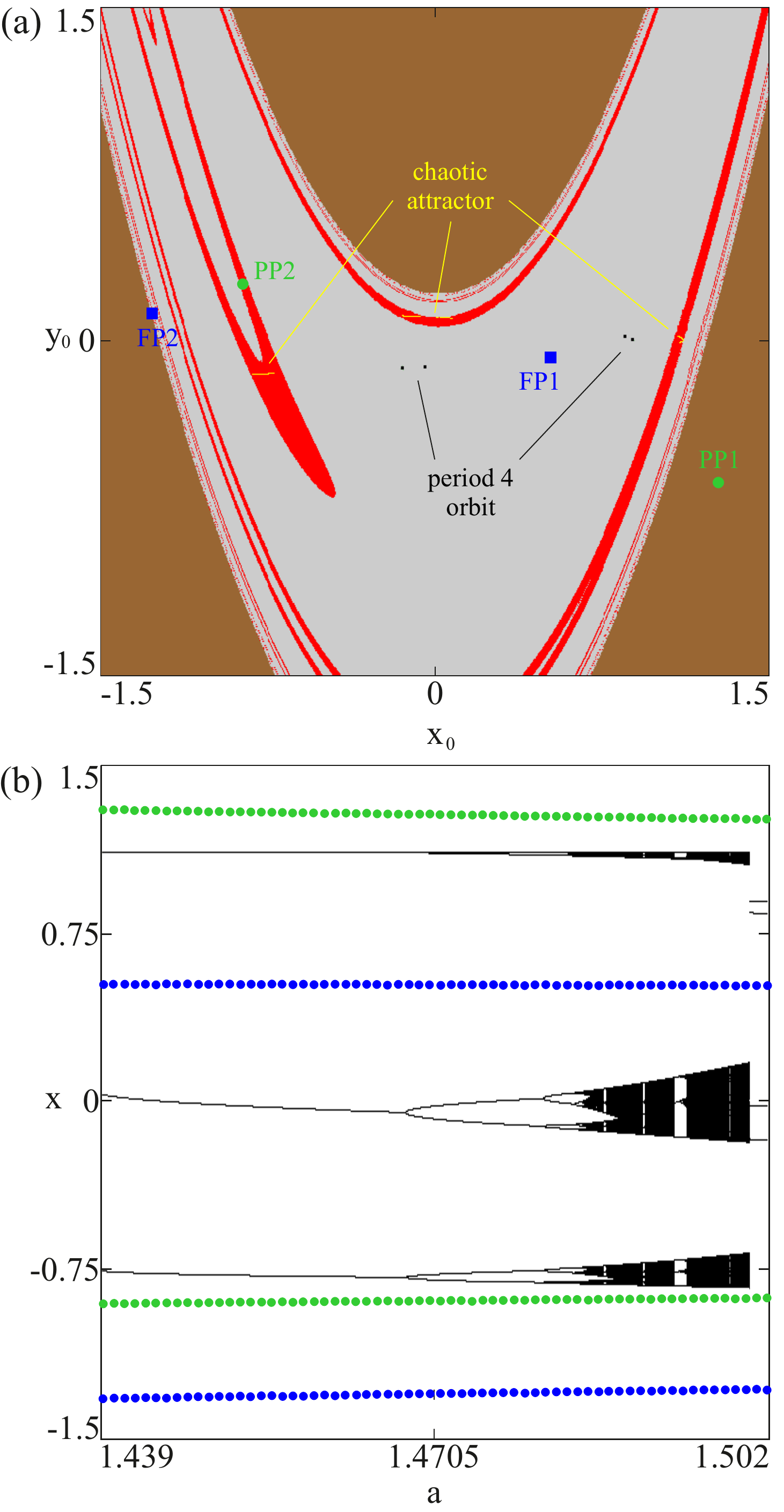}
\caption{(color online). In (a) the basins of attraction of system (\ref{henon}) are presented for $b=0.138$ and $a=1.49$. Period 4 solution (black squares on grey basin) co--exists with hidden chaotic attractor (yellow squares on red basin). Brown region denotes the instability of the system. The bifurcation diagram for $a \in [1.439, 1.502]$ is shown in (b), where period 3 solution has been chosen as the initial state for $a=1.439$. In both subfigures the unstable equilibria and perpetual points are marked in blue and green respectively.}
\label{fig3}
\end{figure}

In our calculations we have fixed parameter $b=0.138$ and analyzed the dynamics and existence of perpetual points for $a \in [1.439, 1.5]$. For such set of parameters system (\ref{henon}) is bi--stable. Typical basins of attraction with corresponding attractors are shown in Fig.~\ref{fig3}(a) for $a=1.49$. The grey basin which attracts the trajectories to the period 4 solution (black squares) co--exists with the red one that leads to the chaotic attractor (yellow squares). The brown area in Fig.~\ref{fig3}(a) denotes the instability region of the map (trajectories escape to infinity).

Map (\ref{henon}) has two unstable fixed points, namely $(x_{fp},y_{fp})=(\frac{b+1+\sqrt{{(b+1)}^2+4a}}{-2a},\frac{-b(b+1+\sqrt{{(b+1)}^2+4a})}{-2a})$ and $(x_{fp},y_{fp})=(\frac{b+1-\sqrt{{(b+1)}^2+4a}}{-2a},\frac{-b(b+1-\sqrt{{(b+1)}^2+4a})}{-2a})$. As can be seen in Fig.~\ref{fig3}(a) (where fixed points are marked by blue squares) one equilibrium FP1=$(0.522, -0.072)$ lies in the grey basin and leads to periodic state, while the second point FP2=$(-1.286, 0.177)$ is located very close to the border between the brown area (infinity) and basins of co--existing attractors. We have examined the initial conditions in the very close neighborhood of the latter critical point and it seems that all the trajectories starting near it either escape to infinity or get attracted by period 4 solution. Hence, the chaotic attractor shown in Fig.~\ref{fig3}(a) by yellow squares is the hidden one \cite{hidden1,hidden3,pp5} (it cannot be located using fixed points of the system).

System (\ref{henon}) for $a=1.49$ has two perpetual points: PP1=$(1.27, -0.631)$ and PP2=$(-0.877, 0.257)$, shown in Fig.~\ref{fig3}(a) as green dots. The first point lies in the brown basin (region of instability), while the second one intersects with the basin of attraction of a hidden chaotic attractor (red basin). Therefore, the attractor can be located using this point.

The bifurcation diagram of the system (\ref{henon}) for parameter $a \in [1.439, 1.502]$ is presented in Fig.~\ref{fig3}(b). As the initial state we have chosen the period 3 attractor (co--existing for $a=1.439$ with period 2 state). The bifurcation scenario is presented as black dots, where on vertical axis the first phase space variable $x$ is shown. The dynamics can be either periodic or chaotic (typical road to chaos through period doubling bifurcations), in contrast to the behavior of co--existing state, which remains only periodic in the whole range $a \in [1.439, 1.5]$. For $a>1.5$ system becomes mono--stable, having one period 4 attractor for $a \in (1.5, 1.502]$. The fixed and perpetual points of the system (\ref{henon}) are shown as blue and green dots respectively.

We have analyzed the attractors presented in Fig.~\ref{fig3}(b) and they remain hidden in the whole range $a \in [1.439, 1.5]$. Moreover, each of them can be located using one of the corresponding perpetual points. Obtained results seems to confirm that perpetual points in maps may be useful in localization of hidden attractors \cite{pp1,pp2,pp5}.
\vspace{\baselineskip}

\textbf{3. Comparison of perpetual points in discrete--time and continuous--time systems}
\vspace{\baselineskip}

(i) Naturally related systems
\vspace{\baselineskip}

In our studies of perpetual points, we have analyzed the systems that have been introduced as maps as well as differential equations.

As the first example, let us consider the continuous-time logistic equation:
\begin{equation}
\dot x = f^{k}_{l}(x),
\label{logODE}
\end{equation}
where $f_l(x)=rx(1-x)$ describes the local dynamics, $r$ is a parameter, and $k \in \mathbb{N}$ denotes the $k$-th superposition of the function $f_l$. The parameter has been considered similarly as for logistic map (\ref{log}), i.e.  $r \in (0,4]$, and for these system (\ref{logODE}) has two fixed points $x=0$ and $x=1$ for every $k$ value. The formula for perpetual points/loci ($\ddot x = 0$) for $k=k^{*}$ is given by:
\begin{equation}
\prod^{k^{*}-1}_{i=1} f^{'}_{l}(\underbrace{f_{l}(\ldots (f_{l}(x)) \ldots)}_{i}) f^{'}_{l}(x) f^{k^{*}}_{l}(x) = 0.
\label{logODEPPs}
\end{equation}
(in the case $k^{*}=1$ product disappear).

\begin{figure}
\includegraphics[scale=0.7]{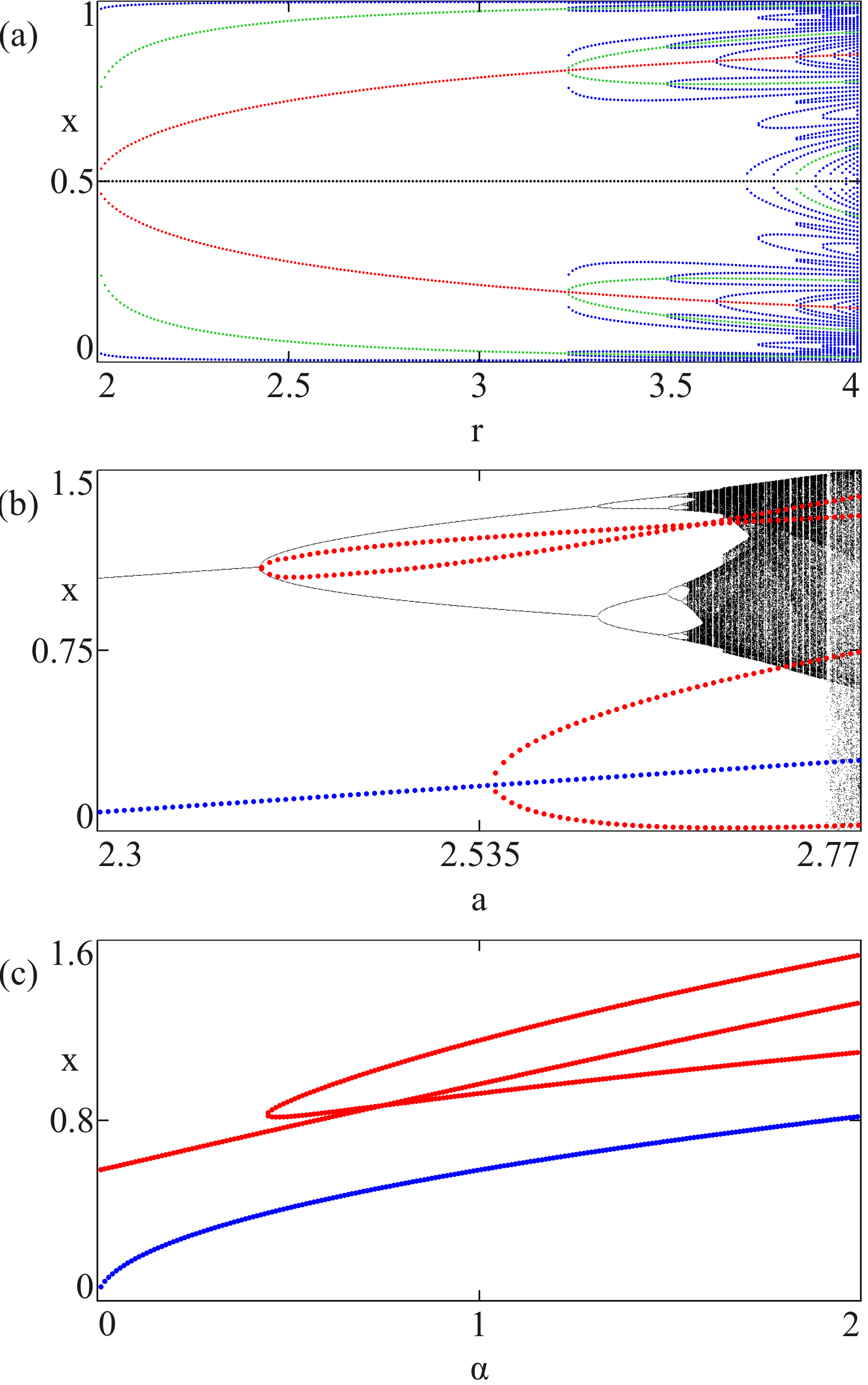}
\caption{(color online). In (a) the perpetual points/loci of system (\ref{logODE}) have been shown for $k=1,2,4,8$ (black, red, green and blue dots respectively). Results obtained for the Duffing map (\ref{DuffingMAP}) and ODEs (\ref{DuffingODE}) are presented in (b) and (c). Standard perpetual points are denoted by blue dots, while period-$2$ perpetual loci are marked by red ones. The bifurcation diagram of map (\ref{DuffingMAP}) is shown in the background of (b) (black dots).}
\label{figDuff}
\end{figure}

Note that the set of solutions of equation (\ref{logODEPPs}) does not include only fixed points of system (\ref{logODE}), but also all perpetual points/loci for $k<k^{*}$. Because of this property, the set of perpetual points/loci for $k=k^{*}$ will be identified as the set of solutions of equation (\ref{logODEPPs}) excluding the points found earlier for $k<k^{*}$.

Our investigations on system (\ref{logODE}) have been shown in Fig.~\ref{figDuff}(a). Perpetual points/loci have been calculated for $k=1,2,4,8$ and are denoted by black, red, green and blue dots respectively. They exist for $r \in [2,4]$ (there are no points for $r<2$) and are located symmetrically around $x=1/2$. It seems that the scenario in which they are born is always the same -- branches of higher $k$ value are born around the solutions of lower $k$, e.g. at $r \approx 3.24$, where two brunches of perpetual loci for $k=4,8$ (green and blue) are created around the red one (for $k=2$). This have been confirmed for each $k=1, \ldots, 8$ (it should be noted that points for some $k$ values are not included in the Fig.~\ref{figDuff}(a)). Additionally, the number of loci increase with the increase of $k$ value (especially for $r>3-3.5$) as well as the parameter $r$ value (this can be observed for $k=4,8$). Therefore the comparison of those results with the ones shown in Figs.~\ref{fig1} and \ref{fig1k} exhibits that dynamics of perpetual points/loci in both cases can be similar.

As the second example, we have considered the Duffing system given in the form of ODEs \cite{DuffODE1,DuffODE2,DuffODE3}:
\begin{eqnarray}
   \left\{
     \begin{array}{l}
     \dot x = g_{1} (x,y) = y,\\
		 \dot y = g_{2} (x,y) = - \beta y + \alpha x - x^3,
     \end{array}
   \right.
	\label{DuffingODE}
\end{eqnarray}

and the Duffing map \cite{DuffMAP1,DuffMAP2,DuffMAP3,DuffMAP4}, i.e.:
\begin{eqnarray}
   \left\{
     \begin{array}{l}
     x_{t+1}= \hat{g}_{1} (x_{t},y_{t}) = y_{t},\\
		 y_{t+1}= \hat{g}_{2} (x_{t},y_{t}) = -b x_{t} + a y_{t} - y^{3}_{t}, 
     \end{array}
   \right.
	\label{DuffingMAP}
\end{eqnarray}
where $\alpha, \beta$ and $a, b$ are the parameters of ODEs and map respectively. The form of equations (\ref{DuffingMAP}) implies from (\ref{DuffingODE}) and finite difference approximation of derivative method \cite{DuffMAP4}.

Based on the concept of \textit{periodic perpetual loci} introduced in Sec. 2(b), analysis of logistic differential equation presented above and standard perpetual points of systems (\ref{DuffingODE}, \ref{DuffingMAP}) we have also analyzed the perpetual loci that result from superpositions of these systems. The period-$2$ perpetual loci have been considered, which result from equations $\dot x = g_{1} (g_{1} (x,y), g_{2} (x,y))$ for continuous--time system and $x_{t+1}= \hat{g}_{1} (\hat{g}_{1} (x_{t},y_{t}), \hat{g}_{2} (x_{t},y_{t}))$ for map (analogous equations for second variables $y$ and $y_{t+1}$).

In Fig.~\ref{figDuff}(b,c) perpetual points/loci ($x$ coordinate) of discrete--time and continuous--time Duffing systems are presented ((b) and (c) respectively). Both systems have a symmetry property (if $(x_{pp}, y_{pp})$ is perpetual, then $(-x_{pp}, -y_{pp})$ is perpetual also), hence only results satisfying $x>0$ have been included. Perpetual points of systems (\ref{DuffingODE}, \ref{DuffingMAP}) are shown as blue dots, while perpetual loci of period-$2$ are presented as red ones. Additionally, in Fig.~\ref{figDuff}(b) the bifurcation diagram of system (\ref{DuffingMAP}) is presented as the background (black dots). As one can see, for considered $a$ and $\alpha$ parameters regions ($b=0.2$ and $\beta=1$ have been fixed) standard perpetual points (which are unique) co--exist with period-$2$ perpetual loci. In the case of map the latter ones are created at period doubling bifurcations and around blue branch of perpetual points. On the other hand, for the Duffing differential equation one can observe two crossing red curves of solutions. The upper one is created around $a \approx 0.45$ and from the very beginning has a form of two branches. As in previous example, the scenarios in which perpetual points/loci are created and their behavior are similar for both types of the Duffing system.
\vspace{\baselineskip}

(ii) Discretization of continuous--time dynamical systems
\vspace{\baselineskip}

Let us consider a general $n$--dimensional continuous--time dynamical system given by:
\begin{equation}
\dot x_i = f_{i} (x_{1}, \ldots, x_{n}),
\label{cont-sys}
\end{equation}
where $i=1, \ldots, n$ and $(x_{1}(t), \ldots, x_{n}(t))$ is the state of the system in time $t \geq 0$.

Typically, system (\ref{cont-sys}) cannot be solved explicitly (except simple linear cases or using complicated analytical procedures), what means that numerical methods have to be applied. The theory of numerical algorithms for differential equations is very well--known \cite{num1,num2,num3}, just to mention standard procedures like the Euler or the Runge--Kutta algorithms \cite{num4}. In our calculations we have used the latter one with order four and typical parameters. Formulas for coefficients are as follows:
\begin{eqnarray}
   \left\{
     \begin{array}{l}
     K^{1}_{i} = f_{i} (x_{1}, \ldots, x_{n}),\\
		 K^{2}_{i} = f_{i} (x_{1}+\frac{1}{2} K^{1}_{1} h, \ldots, x_{n}+\frac{1}{2} K^{1}_{n} h),\\
		 K^{3}_{i} = f_{i} (x_{1}+\frac{1}{2} K^{2}_{1} h, \ldots, x_{n}+\frac{1}{2} K^{2}_{n} h),\\
		 K^{4}_{i} = f_{i} (x_{1}+ K^{3}_{1} h, \ldots, x_{n}+ K^{3}_{n} h),
     \end{array}
   \right.
	\label{coeffs}
\end{eqnarray}
where $h>0$ is the step size.

Using parameters (\ref{coeffs}) system (\ref{cont-sys}) is discretized into $n$--dimensional map of the form:
\begin{equation}
x^{i}_{t+1} = x^{i}_{t} + \frac{1}{6} (K^{1}_{i}+2 K^{2}_{i}+2 K^{3}_{i}+ K^{4}_{i}) h,
\label{discrete-RK4}
\end{equation}
where $i=1, \ldots, n$ and $t \in \mathbb{N} \cup \left\{ 0 \right\}$ is discrete time.

Due to the construction of map (\ref{discrete-RK4}), attractors of systems (\ref{cont-sys}) and (\ref{discrete-RK4}) are naturally related (e.g. both systems have the same equilibria, periodic orbits correspond respectively to the chosen value of step $h$).

Similar considerations can be made if the system (\ref{cont-sys}) is replaced with non--autonomous differential equations.

In order to investigate if the concept of perpetual points in maps is properly defined, and to compare these points in discrete--time and continuous--time systems, we have analyzed some of the systems presented in the original paper about perpetual points \cite{pp1} and their discretized equivalents and calculated the points for both cases.

Systems that have been considered are as follows (values of parameters are included):
\vspace{\baselineskip}

1--dimensional system:
\begin{equation}
\dot x = x^2 - 1
\label{A1}
\end{equation}

2--dimensional system:
\begin{eqnarray}
   \left\{
     \begin{array}{l}
     \dot x = y,\\
		 \dot y = - 2 y + 1.5 x - x^3
     \end{array}
   \right.
	\label{A2}
\end{eqnarray}

3--dimensional system:
\begin{eqnarray}
   \left\{
     \begin{array}{l}
     \dot x = y,\\
		 \dot y = z,\\
		 \dot z = -y + 3 y^2 - x^2 - xz - 0.2
     \end{array}
   \right.
	\label{A3}
\end{eqnarray}

Formulas for perpetual points of systems (\ref{A1} -- \ref{A3}) are given explicitly in \cite{pp1}.

\begin{table}[t]
\centering
    \begin{tabular}{ | l | l | l | l | l |}
    \hline
     & continuous case & discrete case h=0.1 & discrete case h=0.01 & discrete case h=0.001 \\ \hline
    system (\ref{A1}) & 0 & 0.09966777489 & 0.009999666670 & 0.0009999996665 \\ \hline
    system (\ref{A2}) & (-0.7071067811, & (-0.6716265355, & (-0.7035731493, & (-0.7067439618, \\
     & -0.3535533905) & -0.3533782533) & -0.3535526414) & -0.3535516941) \\ \cline{2-5}
     & (0.7071067811, & (0.6716265355, & (0.7035731493, & (0.7067439618, \\
     & 0.3535533905) & 0.3533782533) & 0.3535526414) & 0.3535516941) \\ \hline
    system (\ref{A3}) & (0, & (-0.04741884886, & (-0.004739900755, & (-0.0004738903727, \\
     & 0.4739848152, & 0.4741875881, & 0.4739874456, & 0.4739849853, \\
     & 0) & 0.3757712222 $\cdot 10^{-4}$) & 0.2690515672 $\cdot 10^{-8}$) & -0.1202633626 $\cdot 10^{-9}$) \\ \cline{2-5}
     & (0, & (0.01406694040, & (0.001406375687, & (0.0001406232019, \\
     & -0.1406514819, & -0.1406694062, & -0.1406517878, & -0.1406514392, \\
     & 0) & 0.3287109397 $\cdot 10^{-5}$) & -0.6080132746 $\cdot 10^{-8}$) &  0.5023669888 $\cdot 10^{-10}$) \\ \hline
    \end{tabular}
\caption{In the 2nd column perpetual points of systems (\ref{A1} -- \ref{A3}) are shown, while in the 3rd, 4th and 5th column points of discretized equivalents of these systems are presented (for step size $h=0.1, 0.01, 0.001$ respectively).}
\label{tab}
\end{table}

In Table \ref{tab} perpetual points of systems (\ref{A1} -- \ref{A3}) (2nd column) and their discretized equivalents (columns 3rd-5th) are presented. Calculations for maps have been prepared for different step sizes $h=0.1, 0.01, 0.001$ (3rd, 4th and 5th column of the table respectively). As can be seen, the perpetual points in continuous--time and discrete--time systems are very similar. The difference between these points gets smaller with the decrease of the $h$ value and probably in the limit $h \rightarrow 0$ points overlap. Moreover, the number of perpetual points in both kinds of systems is the same (no new points are found in the map case).

\begin{figure}
\includegraphics[scale=0.7]{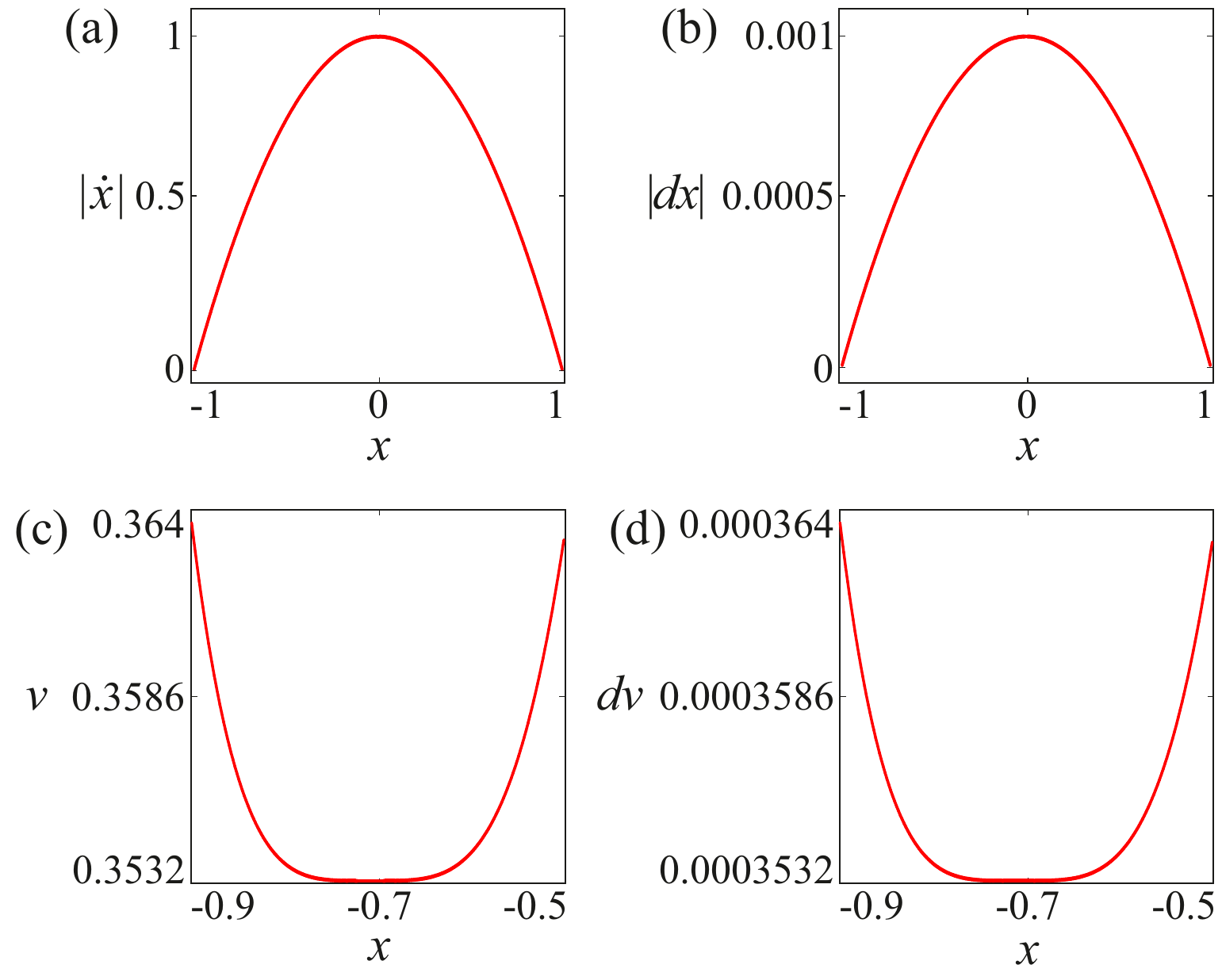}
\caption{(color online). The comparison of velocities near perpetual points between continuous--time systems (\ref{A1}) and (\ref{A2}) ((a) and (c) respectively) and their discretized equivalents as maps (shown in (b) and (d)).}
\label{fig4}
\end{figure}

To investigate the observed similarities we have analyzed the dynamics of systems under consideration near the perpetual points. Results are shown in Fig.~\ref{fig4}. In Figs.~\ref{fig4}(a,b) the absolute velocities of system (\ref{A1}) (left panel) and the discretized map of this system (right panel) are presented (operator $d$ defined in Sec. 2 -- equation (\ref{d-der})). In continuous case the perpetual point is exactly in the maximum $x=0$, while for the map the point is at $x=0.0009999996665$ (calculations have been made for $h=0.001$ step). On the other hand, results obtained for equations (\ref{A2}) are shown in Figs.~\ref{fig4}(c,d). In the left panel one can observe intersection of the magnitude of velocity $v=\sqrt{x^2+y^2}$ near one of the perpetual points (for fixed $y=-0.3535533905$). In the right one analogous intersection of $dv=\sqrt{{(dx)}^2+{(dy)}^2}$ for the map is presented ($y=-0.3535516941$ fixed at perpetual point, $h$ as in previous example). In both cases, the points are near the minimum of the functions.

We have examined other examples of continuous--time systems and their corresponding maps and the similar shape of velocities, like the one shown in Fig.~\ref{fig4} seems to be a universal phenomenon, provided the step used in the discretization is sufficiently small. Also, the velocity is proportional to its discrete equivalent with the coefficient $1/h$ (e.g. in the case of one--dimensional system $\dot x \approx \frac{1}{h} \cdot d x$). The latter fact arise from the construction of \textit{discrete derivative} (\ref{d-der}) and equation (\ref{discrete-RK4}) in which component $\frac{1}{6} (K^{1}_{i}+2 K^{2}_{i}+2 K^{3}_{i}+ K^{4}_{i})$ describes the velocity field.

Additionally, in our research we have compared the acceleration and \textit{second order discrete derivative} (continuous-- and discrete--time systems respectively) and the results we got also show similar properties for both structures. Functions are proportional, although in this case the proportionality coefficient equals $1/{h}^2$ (in the case of one--dimensional system $\ddot x \approx \frac{1}{h^2} \cdot d^2 x$).
\vspace{\baselineskip}

\textbf{4. Conclusions}
\vspace{\baselineskip}

In this paper, we have shown that the concept of perpetual points, originally introduced in continuous--time systems can be successfully transformed to maps. Research on perpetual points and periodic perpetual loci can improve our understanding of the system and their analysis may be essential in studying the system dynamics. The calculations can be made more easily than for ODEs due to the fact that maps are usually less complicated than flows. Presented examples confirm that this kind of points/loci is common in discrete--time systems, and can be found using standard numerical methods.

The analysis of perpetual points/loci in systems that have representatives in maps and ODEs shows that the behavior of these states and their occurrence is similar for both cases. This have been confirmed in most typical examples (Sec. 3(i)). Obtained results are useful in studies on connections between perpetual points/loci in different types of systems. Nevertheless, to better understand these relations, more complex systems have to be investigated. 

On the other hand, the connection in the case of discretization of ODEs is more clear. If we transform the continuous--time system into a map using one of the standard numerical algorithms, the values of points obtained in both cases will be very close and the dynamics across these points will be similar.

All presented similarities may confirm that the definition of \textit{discrete derivative} (equations (\ref{d-der}) and (\ref{dd-der})) and resulting perpetual points/loci are properly defined and can be a useful tool in studies of maps dynamics.

Perpetual points in maps may have similar properties, as in continuous--time systems. Especially, they may provide a guidance to hidden attractors, like presented for Henon map example (equation (\ref{henon})). Nonetheless, this connection requires a deeper analysis.

The concept we have introduced in this paper seems to be universal and similar results can be observed in other systems with different dynamics. However, the character of perpetual points/loci in dynamical systems is still not sufficiently well understood and further research need to be established, in continuous--time as well as in discrete--time systems.
\vspace{\baselineskip}

{\bf Acknowledgment}
This work has been supported by the Polish National Science Centre, MAESTRO Programme -- Project No 2013/08/A/ST8/00/780.
AP  acknowledges the DST, Govt. of India for financial support, and 
thanks the TU, Lodz, for warm  hospitality during several visits.


\begin{thebibliography}{60}

\bibitem{pp1}
A. Prasad, Int. J. Bifurcation Chaos 25, 2, 1530005 (2015).

\bibitem{pp3}
A. Prasad, A Note On Topological Conjugacy For Perpetual Points, I. J. Non. Sc. (2016) [http://arxiv.org/abs/1511.05836 (2015)]

\bibitem{pp2}
D. Dudkowski, A. Prasad, T. Kapitaniak, Phys. Lett. A 379, 40-41, 2591-2596 (2015).

\bibitem{pp6}
T. Ueta, D. Ito, K. Aihara, Int. J. Bifurcation Chaos 25, 13, 1550185 (2015).

\bibitem{pp4}
S. Jafari, F. Nazarimehr, J. C. Sprott, S. M. R. H. Golpayegani, Int. J. Bifurcation Chaos 25, 13, 1550182 (2015).

\bibitem{pp5}
D. Dudkowski, S. Jafari, T. Kapitaniak, N. V. Kuznetsov, G. A. Leonov, A. Prasad, Phys. Rep. 637, 1 (2016).

\bibitem{diss}
G. A. Leonov, N. V. Kuznetsov, T. N. Mokaev, Eur. Phys. J. Special Topics 224, 1421-1458 (2015). 

\bibitem{may}
R. M. May, Nature 261, 459-467 (1976). 

\bibitem{fin}
A. J. G. Cairns,  Interest Rate Models - An introduction, Princeton University Press (2004).

\bibitem{radio}
E. Salinelli F. Tomarelli, Discrete Dynamical Models, Springer (2014).

\bibitem{dose}
F. R. Adler, Modeling the Dynamics of Life: Calculus and Probability for Life Scientists, Cengage Learning (2012).

\bibitem{bearing1}
Y. Isagi, K. Sugimura, A. Sumida, H. Ito, J. Theor. Biol. 187, 231-239 (1997). 

\bibitem{bearing2}
A. Prasad,  K. Sakai, Chaos 25, 123102 (2015).

\bibitem{maps2}
L. Zhang, IEEE Trans. Circuits Syst. I, Reg. Papers 58, 5, 1109-1118 (2011).

\bibitem{maps4}
K. You, IEEE Trans. Autom. Control 56, 10, 2262-2275 (2011).

\bibitem{maps3}
F. Amato, M. Ariola, C. Cosentino, Automatica 46, 5, 919-924 (2010).

\bibitem{hidden4}
G. A. Leonov, N. V. Kuznetsov, V. I. Vagaitsev, Phys. Lett. A 375, 23, 2230-2233 (2011).

\bibitem{hidden5}
H. Zhao, Y. Lin, Y. Dai, Int. J. Bifurcation Chaos 24, 6, 1450080 (2014).

\bibitem{hidden6}
A. P. Kuznetsov, S. P. Kuznetsov, E. Mosekilde, N. V. Stankevich, J. Phys. A: Math. Theor. 48, 12, 125101 (2015).

\bibitem{hidden8}
V.-T. Pham, S. Vaidyanathan, C. K. Volos, S. Jafari, Eur. Phys. J. Special Topics 224, 1507-1517 (2015).

\bibitem{hidden7}
U. Chaudhuri, A. Prasad, Phys. Lett. A 378, 9, 713-718 (2014).

\bibitem{hidden1}
G. A. Leonov, N. V. Kuznetsov, Int. J. Bifurcation Chaos 23, 1, 1330002 (2013).

\bibitem{hidden3}
G. A. Leonov, N. V. Kuznetsov, O. A. Kuznetsova, S. M. Seledzhi, V. I. Vagaitsev, Trans. Syst. Control 2, 6 (2011).

\bibitem{hiddenmaps1}
R. Alli-Oke, J. Carrasco, W. Heath, A. Lanzon, A robust Kalman conjecture for first-order plants, Proc. IEEE Control and Decision Conference (2012).

\bibitem{hiddenmaps2}
W. P. Heath, J. Carrasco, M. de la Sen, Automatica 60, 140-144 (2015).

\bibitem{hiddenmaps3}
S. Jafari, V.-T. Pham, S. Golpayegani, M. Moghtadaei, S. T. Kingni, The relationship between chaotic maps and some chaotic systems with hidden attractors, Int. J. Bifurcation Chaos (2016).

\bibitem{hiddenmaps4}
H. Jiang, Y. Liu, Z. Wei, and L. Zhang, Hidden chaotic attractors in a class of two-dimensional maps, Nonlinear Dynam. [published online] (2016).

%

\bibitem{log1}
M. Ausloos, M. Dirickx, The Logistic Map and the Route to Chaos, Springer (2006).

\bibitem{log2}
L. Kocarev, G. Jakimoski, Phys. Lett. A 289, 4-5, 199-206 (2001).

\bibitem{log3}
N. K. Pareek, V. Patidar, K. K. Sud, Image Vision Comput. 24, 9, 926-934 (2006).

\bibitem{scaling}
Y. Saiki, M. Yamada, Nonlin. Processes Geophys. 15, 675-680 (2008).

\bibitem{tent2}
Y. Maistrenko, T. Kapitaniak, Phys. Rev. E 54, 3285 (1996).

\bibitem{tent3}
D. Dudkowski, Y. Maistrenko, T. Kapitaniak, Phys. Rev. E 90, 032920 (2014).

\bibitem{tent1}
T. Yoshida, H. Mori, H. Shigematsu, J. Stat. Phys. 31, 2, 279-308 (1983).

\bibitem{henon1}
P. Grassberger, H. Kantz, U. Moenig, J. Phys. A: Math. Gen. 22, 5217-5230 (1989).

\bibitem{henon2}
G. D'Alessandro, P. Grassberger, S. Isola, A. Politi, J. Phys. A: Math. Gen. 23, 5285-5294 (1990).

\bibitem{henon3}
O. Biham, W. Wenzel, Phys. Rev. A 42, 8, 4639-4646 (1990).

\bibitem{henon4}
P. Kuzma, Dynamics of the coupled Henon maps, Master of Science Thesis (2012).

\bibitem{DuffODE1}
E. Ott, Chaos in Dynamical Systems, Cambridge University Press (1993).

\bibitem{DuffODE2}
I. Kovacic, M. J. Brennan, The Duffing Equation: Nonlinear Oscillators and their Behaviour, Wiley (2011).

\bibitem{DuffODE3}
U. Parlitz, W. Lauterborn, Phys. Lett. A 107, 8, 351-355 (1985).

\bibitem{DuffMAP1}
P. G. Reinhall, T. K. Caughey, D. W. Storti, J. Appl. Mech 56, 1, 162-167 (1989).

\bibitem{DuffMAP2}
O. Junge, Uncertainty in the dynamics of conservative maps, 43rd IEEE Conference on Decision and Control (2004).

\bibitem{DuffMAP3}
I. M. T. AL-Shara'a, M. A. A.-K. AL-Yaseen, Mathematical Theory and Modeling 3, 7, 41-45 (2013).

\bibitem{DuffMAP4}
\url{http://www.cmp.caltech.edu/~mcc/Chaos_Course/Lesson5/Maps2D.pdf}

\bibitem{num1}
G. Hall, J. M. Watt, Modern Numerical Methods for Ordinary Differential Equations, Clarendon Press (1976).

\bibitem{num2}
C. W. Gear, L. R. Petzold, SIAM J. Numer. Anal. 21, 4, 716-728 (1984).

\bibitem{num3}
E. Hairer, C. Lubich, G. Wanner, Geometric Numerical Integration: Structure-Preserving Algorithms for Ordinary Differential Equations, Springer (2010).

\bibitem{num4}
J. C. Butcher, The numerical analysis of ordinary differential equations: Runge-Kutta and general linear methods, Wiley-Interscience New York (1987).

\end{thebibliography}
\end{document}